\documentclass[aps, prl, reprint]{revtex4-1}

\usepackage{amssymb, amsmath, amsthm, xcolor}
\usepackage{mathtools}

\newcommand{\bn}{\boldsymbol n}
\def\rhoDNA{\rho_{\begin{tiny}{\textrm{DNA}}\end{tiny}}}
\def\MDNA{M_{\begin{tiny}{\textrm{DNA}}\end{tiny}}}

\def\fcr{F_{{\textrm{\begin{tiny}{Chr}\end{tiny}}}}}
\def\frel{F_{{\textrm{\begin{tiny}{Rlx}\end{tiny}}}}}

\def\fn{F_{{\text{\begin{tiny}{N}\end{tiny}}}}}

\newcommand{\partderiv}[2]{\ensuremath{\frac{\partial #1}{\partial #2}}}

\def\bm{\boldsymbol m}
\def\bp{\boldsymbol p}
\newcommand{\R}{\mathbb{R}}
\DeclareMathOperator{\tr}{tr}
\def\bx{\boldsymbol  x}
\def\br{\boldsymbol  r}
\begin{document}
\title{A Liquid Crystal Model of  Viral DNA Encapsidation }
\author{Javier Arsuaga, Maria Carme Calderer, Lindsey Hiltner and Mariel V\'azquez}
\date{\today}
 \protect\thanks{The authors  acknowledge the discussions and advice from Dr.~Oleg Lavrentovich. Arsuaga, Calderer and Vazquez wish also to acknowledge the support from the grants DMS/NIGMS R01 GM109457, NFS-DMREF 1435372 and DMS1057284, respectively.{\tiny }  }

\begin{abstract}
A liquid crystal continuum modeling framework for icosahedra bacteriophage viruses is developed and tested. The main assumptions of the model are the chromonic columnar hexagonal structure of confined DNA, the high resistance to bending and the phase transition from solid 
to fluid-like states as the concentration of DNA in the capsid decreases during infection. The model predicts osmotic pressure inside the capsid and the ejection force of the DNA as well as the size of the isotropic volume at the center of the capsid. Extensions of the model are discussed. 
\end{abstract}

\maketitle

Double-stranded (ds)DNA bacteriophages are of renewed interest due to their use in medicine \cite{haq2012bacteriophages,Sulakvelidze2001} and biotechnology \cite{haq2012bacteriophages}. Icosahedral bacteriophages consist of a protein capsid with icosahedral symmetry whose assembly is followed by the packing, by a molecular motor, of a single naked dsDNA molecule  \cite{Smith2001}.  The DNA molecule inside the viral capsid is found under extreme concentration and osmotic pressure. At the time of infection the DNA is released by a mechanism that suggests a phase transition, possibly into a 'liquid-like' state \cite{Leforestier2010, Liu2014solid, Sae2014}. Both processes, packing and releasing of the genome, are highly dependent on how the DNA molecule folds inside the viral capsid; however our understanding of this folding remains very limited. 

%Therefore there are three critical and interrelated aspects that need to be properly characterized for the design of bacteriophages as delivery machines: the packing reaction, the folding of the viral genome inside the capsid, and the ejection reaction of the viral genome at the time of infection. 

 The concentration of the DNA molecule inside the viral capsid is between $200$ and $800$ $mg/ml$ \cite{Kellenberger1986} and the estimated osmotic pressure ranges between 40 and 60 atmospheres \cite{Evilevitch2003,Jeembaeva2008}. Three factors contribute to the excess pressure found inside the viral capsid:  the decrease in entropy associated with the confinement imposed by the capsid, the high resistance of the DNA molecule to bending beyond its persistence length and the self-repulsion of the DNA molecule \cite{RiemerBloomfield1978}. Experimental and theoretical studies acquired over the last 30 years  \cite{Leforestier2009,Leforestier2010,Lepault1987,Reith2012,Rill1986,Strzelecka1988, park2008self} have shown that  under such conditions the DNA molecule forms a columnar  hexagonal liquid crystal phase. In particular, liquid crystalline phases in bacteriophages were first proposed in \cite{Kellenberger1986}, with an explicit reference to 
hexagonal packing made in \cite{Livolant1991}
 and since then, consistent data have been accumulating \cite{Leforestier1993,leforestier2008bacteriophage,Marenduzzo2009,Reith2012}. %Moreover, in conditions of isolation such as inside a capsid, the packing structure is that of a solid crystal, allowing for very limited DNA mobility. %including hexagonal chromonics [CARME]. 

A number of theoretical models, based on experimental and/or simulation data, have been proposed to describe the folding of the DNA molecule inside the bacteriophage capsid (e.g.\cite{Cerritelli1997,Earnshaw1980,Hud1995,Leforestier2010,Lepault1987,Petrov2007a,Serwer1992}) with only few attempts to describe the DNA molecule in a liquid crystalline phase \cite{Arsuaga2008,Marenduzzo2009}. These attempts however were based on energy fields originally proposed for modeling  DNA molecules in free solution \cite{Arsuaga2002a,Arsuaga2002b,Arsuaga2005,Marenduzzo2013,Marenduzzo2009,Petrov2007a,Reith2012,Rollins2008,Spakowitz2005} and do not provide adequate description of DNA inside the phage capsid. 

%%%%%%%%%%%%%%%%%%%%%%%%%%%%%%%
In this article, we use cryo-electron microscopy (cryo-EM) data, the hexagonal chromonic liquid crystal structure of the packed DNA \cite{Livolant1991}, and the continuum theory of liquid crystals to build a new model of DNA folding inside the viral capsid. The final packed structure of the DNA molecule corresponds to an energy minimizing configuration of the energy proposed below. 

\begin{figure}
  	\centering
                      \includegraphics[width=\linewidth]{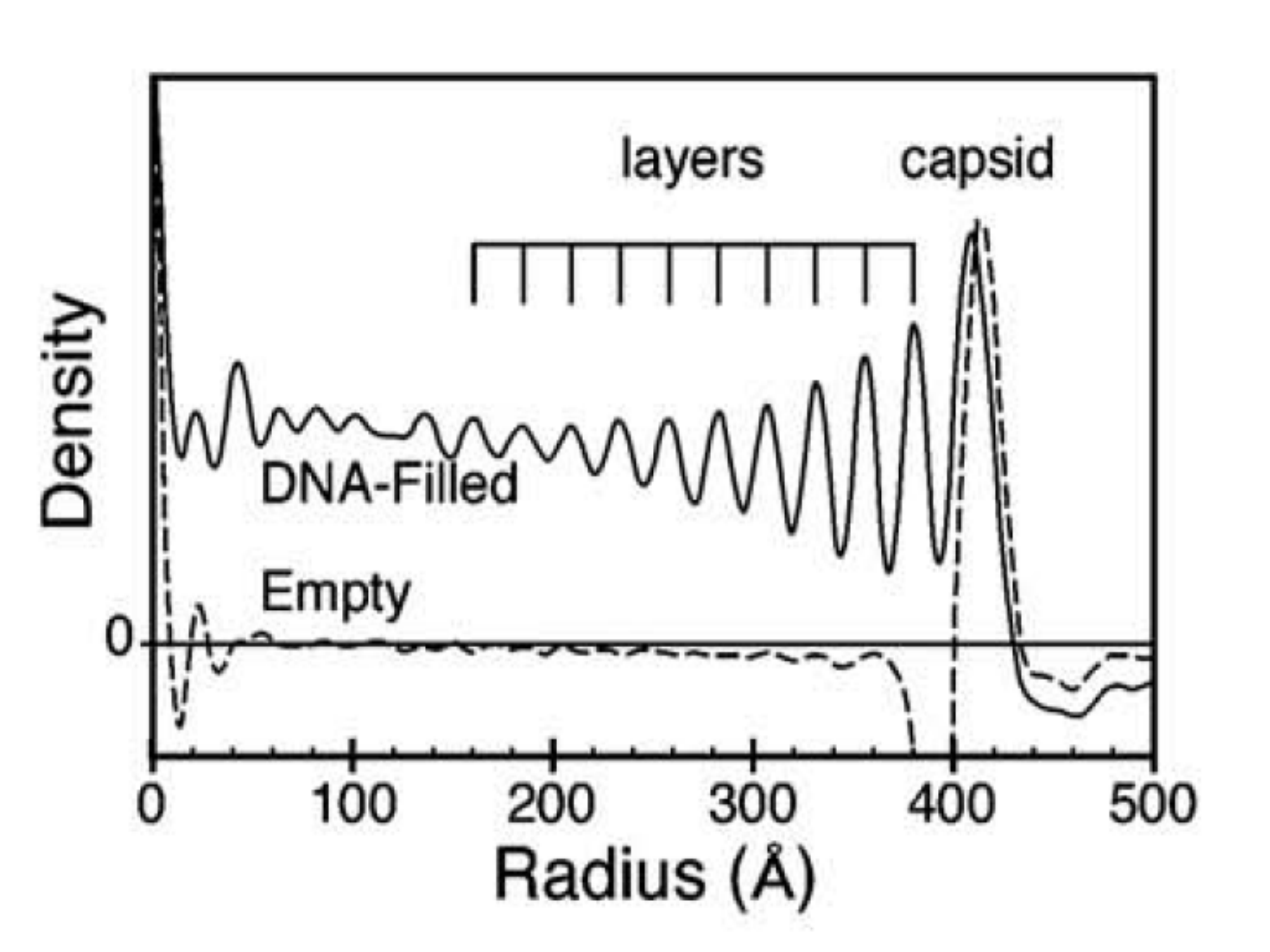}
        \caption{DNA density of bacteriophage $T5$. The graph shows the radial density distribution of DNA inside the viral capsid. The dashed line shows the density of an empty capsid. Figure reproduced from \cite{Molineux2013}}\label{EM}
   \end{figure}

Our model assumes that the hydrated DNA fills the entire volume of the capsid \cite{kellenberger1986considerations} hence two relevant parameters are the molar $c$ and volume $c_v$ concentration of DNA. We take the customary point of view that 
 the DNA molecule is a semiflexible elastic polymer characterized by the persistence length $L_p$ which is about the size of the radius of the capsid. 
Since Cryo-EM data for most bacteriophages present multilayered spooling-like configurations on the outer layers of the packed genome \cite{Earnshaw1980,Cerritelli1997,Comolli2008,Chang2006,Effantin2006,Leforestier2009,Jiang2006} we assume that the hexagonal chromonic structure provides a geometric scaffolding that sustains the trajectory of the DNA molecule. The decreasing available volume during DNA packing induces an extreme bending on the DNA molecule whose bending resistance prevents the hexagonal ordering from completely filling the capsid, so an inner core of disordered/isotropic DNA is assumed to form (See Figure \ref{EM}) \cite{Comolli2008,Chang2006,Effantin2006,Jiang2006}. CryoEM data also suggest an axisymmetric capsid with parallel and meridian arrangements of the DNA molecule.    
  %Indeed, the model has the ability to predict the  pressure inside the capsid.
Data at our disposal include: (1) cryoelectron images of the bacteriophage exhibiting capsid shape and protein core, the size and shape of the disordered region, the volume $V$ of the capsid and the DNA density graphs that allow us to obtain the number, $M$, of concentric layers \cite{Comolli2008,Chang2006,Effantin2006,Jiang2006}; (2) ordering of the DNA molecule at the boundary as promoted by the capsid \cite{Earnshaw1980,Effantin2006,Lepault1987,Cerritelli1997};  (3) DNA effective diameter $d$ and genome length $L$;  (4) pressure measurements as well as speeds of DNA ejection \cite{cordova2003osmotic, Evilevitch2003, Grayson2006, Leforestier2009, Tzlil2003}.

To make our model precise, we assume that the capsid corresponds to a bounded, discretely axisymmetric region
 $\mathcal{B}\in \R^3$ with  $\partial \mathcal B$
 %$\mathcal S_0$
  representing the piecewise smooth, faceted, viral capsid. We let $l_0>0$ denote the length of the axis that connects the location of the connector with its antipodal site in the viral capsid. 
%Let $O\in \mathcal{B}$ represent the center of the capsid, that we also take as the origin of the coordinate system and $\mathbf{x}\in\mathcal B$ a point in the capsid. 
Let $\Omega_0\subset\mathcal B$ denote the isotropic region of the capsid, also taken to be axisymmetric with respect to $l_0$ and $\Omega:=\mathcal B \setminus\Omega_0$, nonempty, denote the region occupied by the hexagonal chromonic liquid crystal phase. A piecewise smooth curve $\br=\br(s), s\in [0,L]$, describes the axis of the DNA. Note that, in general, the size of $ {\Omega}_0$ is an unknown of the problem. 

%
%The role of the capsid proteins in promoting filament ordering motivates the choice of boundary conditions of the problem. We first identify the capsid as the level surface 
%${\mathcal S}_{\mathbf{m}}^0$ of model, and consider the family of smectic layer curves ${\mathcal C}_{i,0} \in {\mathcal S}_{\mathbf{m}}^0$, and let $\mathbf{T}, \mathbf{n} $ and $\mathbf{B}$ represent the Fr{\'e}net-Serret vectors on a point of the curve. We require that, for $\mathbf{x}\in {\mathcal C}_{i,0}$,
%\begin{eqnarray}
%&&\mathbf{n} = \mathbf{T}, \mathbf{m} = -\boldsymbol N, \, \mathbf{p} = \mathbf{B}, \frac{\partial\omega}{\partial\boldsymbol N}= q, \frac{\partial\vartheta}{\partial\mathbf{B}}= q. \label{bc0}\end{eqnarray}}

The model of hexagonal chromonic liquid crystals that we present  corresponds to that of de Gennes in the case of small distortions (\cite{dGP93}, sec.7.1).  A triple of linearly independent unit vectors $\bn, \bm, \bp$ represents the uniaxial nematic director, along the direction tangent to the axis of the DNA molecule,  and   two local directions of ordering, respectively. The vectors $\bm$ and $\bp$ correspond to the lattice vectors of the columnar phase and account for the meridian and parallel arrangements. We will assume that the  three vectors are almost mutually perpendicular in the sense made precise by the energy described below.
 In an energy minimizing configuration, these directions will be determined from their corresponding boundary values, that is, the filament organization in the contact with the capsid. Two 
 complex valued functions $\psi= \rho e^{iq\omega}$ and $ \gamma=\rho e^{i q\vartheta}$ account for the density of ordered material and  describe the space filling conformation. Specifically, $\rho\geq 0$ gives the packing density of the layered molecules ($\rho=0$ corresponds to disordered DNA) along the two preferred directions.
 The real number $q$ 
 corresponds to the frequency of the layers with  $d = 2\pi/q$ representing the the effective diameter of the DNA and interlayer distance. 

\begin{figure}
  	\centering
                      \includegraphics[width=\linewidth]{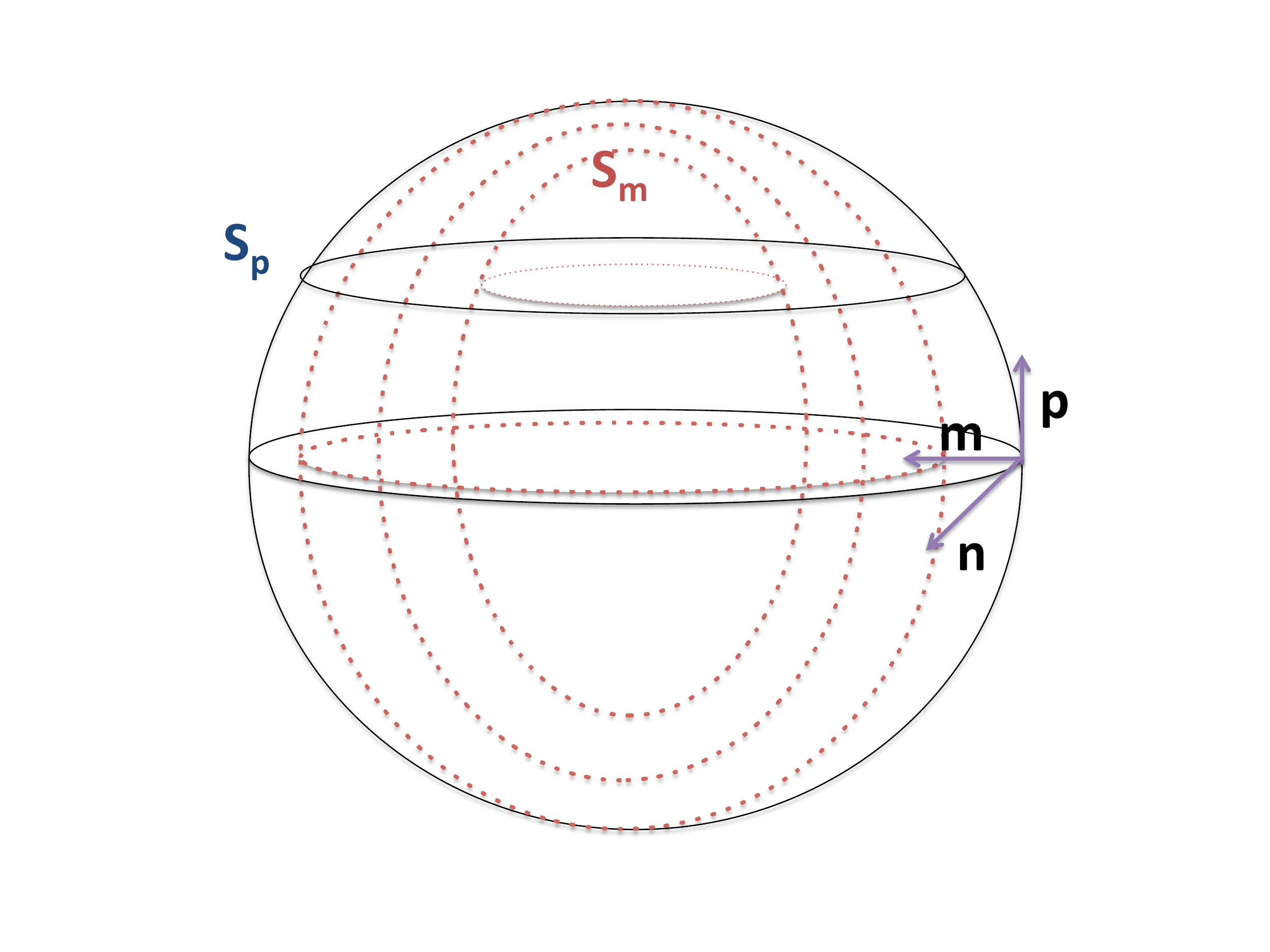}
        \caption{This is a schematic representation of the level sufaces $\omega$=constant, \,$\vartheta$= constant, whose intersections provide the scaffolding that supports the DNA filament.}\label{LevelSurfaces}
   \end{figure}

 The chromonic structure  is described by a discrete  2-family of level surfaces (see Figure \ref{LevelSurfaces})
\begin{align}
&{\{\mathcal S_{\bm}^i\}}_{i=0}^M: \, \omega(\bx)=id,\quad 
{\{\mathcal S_{\bp}^j\}}_{j=0}^P: \, \vartheta(\bx)=jd\label{level-surfaces}\end{align}
where $M$ and $P=[\frac{l_0}{d}]$ (with brackets indicating the integer immediately below the value of the quotient) are nonzero positive integers. The first family corresponds to spheres concentric with the capsid while the second are the planes defined by $z=$constant.    
 As part of satisfying the boundary conditions of the problem, we take $S_{\bm}^0\equiv \partial \mathcal B$, with $j=0$ corresponding to the parallel surface $z=-\frac{l_0}{2}$. These choices correspond to a spooling direction starting on the lower part of the capsid surface.
The family of  intersecting curves $\mathcal C_{i,j} = \mathcal{S}_{\mathbf{m}}^i\cap\mathcal{S}_\mathbf{p}^j, \, 0\leq i\leq M, \, 0\leq j\leq P,$ give the location of the axis of the DNA molecule in the capsid and its ordering determines the direction of DNA spooling in the capsid. The final outcome, however, is independent of the ordering choices.

%Each one represents a 'reel' surface that will span the filament. 

 We propose that the total energy is given by the sum of the ordered chromonic phase plus the disordered isotropic core $\Omega_0$ and a surface energy penalizing the interface between the two:
\begin{equation}
E =\int_{\Omega}\fcr\,d\bx + \nu\textrm{Vol}(\Omega_0)+\sigma \textrm{Area}(\partial \Omega_0), \,\, \textrm {with} \label{total-energy}
\end{equation}
 \begin{widetext}
 	\begin{align}
 	\fcr&=\fn(\nabla\mathbf{n}, \mathbf{n}) + \fn(\nabla\mathbf{m}, \mathbf{m}) + \fn(\nabla\mathbf{p}, \mathbf{p})+ C_d\big(|\nabla\psi-iq\psi\mathbf{m}|^2 + |\nabla\gamma-iq\gamma\mathbf{p}|^2\big)+\frel(\bn, \bp, \bm)
 	\label {energy-na}\\
\fn(\nabla\bn, \bn)& = k_1{(\nabla\cdot\bn)}^2+k_2{(\bn\cdot\nabla\times\bn)}^2+ k_3|\bn\times(\nabla\times\bn)|^2 +(k_2+k_4)(\tr{(\nabla\bn)}^2-(\nabla\cdot\bn)^2)\label{oseen-frank}\\
\frel(\bn, \bm,\bp)&=  A (\mathbf{m}\cdot\mathbf{n})^2+B (\mathbf{m}\cdot\mathbf{p})^2+ C(\mathbf{n}\cdot\mathbf{p})^2.\label{f-relax} 
 	\end{align}
\end{widetext}
%% Revise the geometric interpretation of the O-F energy associated with m and p. Would a gradient term be sufficient?
%% It may be convenient to single out the role of the  OF energy for n, associating it with the energy of the filament. Maddocks.
%Possible optimization approach to obtain parameters 'machine learning'.  Introduce the idea of subsequent model improvement using the virus data bank. Portray our model as a prototype. 
Not all the terms in the energy are relevant to the continuum theory. Assuming that $k_1$ and $k_2$ diverge near the phase transition to the hexagonal phase leads to a pure bending distortion at the limit (\cite{dGP93}, sect 7.1; \cite{kleman2007soft}, page 315). 
The  bending modulus is taken to be that of a semiflexible polymer in confinement,  $k_3= K_BTL_p I^{-1}$. I is the geometric moment of inertia with respect to the capsid axis. The value of I depends on 
 the distance of the DNA molecule to to the capsid axis, and so, it accounts for the increase in bending resistance as the axis $l_0$ is approached.
  For simplicity, we associate the Oseen-Frank energy function
 $F_N(\nabla\bn, \bn)$ to the director fields $\bm$ and $\bp$, expressing, in particular, high resistance to bending, to ensure a solid-like packing in the ordered region. This corresponds to the assumption that the filament properties also determine the geometry of packing. 
  As for lyotropic liquid crystals, the energy density $\nu$ of the isotropic phase is assumed to be a  function of the molar concentration $c$. Specifically,  we use the approximate expressions derived by   Onsager for  lyotropic liquid crystals, that in large concentration regimes  \cite{onsager1949} take the form:
$\nu=K_B T\ln(c^2-\frac{45}{8}+ c^{-2})-1$. The surface energy density $\sigma = K_B T\frac{0.257}{L_p d}$ is based on the Onsager's theory and derived in \cite{doi-kuzuu1985} for such regimes. (Approximations appropriate to  low concentration regimes can also be  found in the aforementioned references). The quantity $C_d$ corresponds to the compressibility modulus.  The volume concentration $c_v$, which measures how much of the capsid volume is occupied by DNA, together with  the observation that the genome tends to fill the entire capsid \cite{purohit2005forces} leads us to assume that $C_d=C_d(c_v)$ satisfying  $ C_d(0)=0,$  $C_d'(c_v)>0$ and 
$\lim_{c\to 1^{-}}C_d(c_v)=+\infty, $ the latter enforcing an idealized perfect packing structure at the limit of high concentrations. Note that these assumptions on $C_d$ encode the phase transition solid to liquid-like behavior since the contribution of the chromonic phase will decrease as $c_v$ decreases at the time of infection. 
%\begin{eqnarray}
%&&\mathbf{n} = \mathbf{T}, \mathbf{m} = -\boldsymbol N, \, \mathbf{p} = \mathbf{B}, \frac{\partial\omega}{\partial\boldsymbol N}= q, \frac{\partial\vartheta}{\partial\mathbf{B}}= q. \label{bc0}\end{eqnarray}}
The energy  (\ref{f-relax}) allows for the relaxation of the orthogonality constraints between pairs of vectors $\bn, \bm, \bp$. 
The positive constants $A, B, C$ are then taken to be larger than the maximum dimensionless parameter of the energy. The combination of the energy terms (\ref{oseen-frank}) and (\ref{f-relax}) is analogous to the Ginzburg-Landau energy 
governing many condensed matter processes \cite{Landau1950}. 

The role of the capsid proteins in promoting ordering of the DNA molecule
dictates the choice of boundary conditions of the problem. If $\bm_0$ denotes the normal vector to $\partial\mathcal B$ at a point $\bx \in \mathcal B$ (where the normal is well defined) and $\bn_0$ the unit tangent to the axis of the DNA at $\bx$,  we take $\bp_0:=\bn_0\times\bm_0$;  note that $\bn_0, \bm_0 $ and $\bp_0$ form the local Fr{\'e}net-Serret system associated with the curve through $\bx$ and serve as Dirichlet boundary conditions for the unknown fields $\bn, \bm$ and $\bp$.  In addition, we require $\partderiv{\omega}{\bm}=q$ and $\partderiv{\vartheta}{\bp}= q $ at $\bx$; this corresponds to an idealized perfect packing on the capsid surface. 

In the case  that the disordered  core $\Omega_0$ is a fixed region and the Frank constants, $k_i$, satisfy coercivity inequalities, standard methods of calculus of variations yield existence of a minimizer of the energy subject to the previously stated boundary conditions \cite{HaKiLi86}. This configuration is piecewise smooth 
except for a discrete collection of singular points where $\bn=0$.
These correspond to defects of the Oseen-Frank theory. Consequently,
the level curves $\mathcal C_{ij}$ are piecewise smooth. 

The reconstitution of the trajectory of the dsDNA molecule is achieved by subsequently solving the  initial value problem for the ordinary differential equation $\br'(s)= \bn(\br(s))$, $|\bn|=1$,  $s\in [0,l]$ with $r(\mathbf 0)=\br_0$, where $\br_0$ represents the location  
of the attached filament tip at the entrance of the capsid. 
The solution curve  $\br(s)$ describing the trajectory of the DNA axis is piecewise differentiable and   provides the parametrization of the curves $\mathcal C_{i,j}$; it  becomes singular at the liquid crystal defect locations $\bn=\boldsymbol 0$.
The latter are also the sites where the filament transitions from the curve segment $\mathcal C_{i,j}$ to either 
 $\mathcal C_{i,j+1}$ or $\mathcal C_{i+1,j}$, following  the direction of spooling. 
 
 %This ordering also endows the curve $\br(s)$ with  orientation.  (Already mentioned above)
 
% Defects are  also associated with locations of kinks in the DNA filament. 
% As the DNA spools in the capsid, transitions from one segment $\mathcal C_{i, j}$ to the next, 
% $\mathcal C_{i+1, j} $ or $\mathcal C_{i, j+1}$, take place at defect locations. 
Filament crossings within a curve segment $\mathcal C_{i,j}$ also correspond to liquid crystal defects. 
 These include: (a) dislocations, points with $\rho = 0$,  where two or more layers of the same family come into contact allowing the filament to break the order of transition from $\mathcal C_{i, j}$ to $\mathcal C_{i+1, j} $; 
 (b) point defects where the curve becomes singular and filament contacts may take place to perform a crossing; (c) point defects of order $\pm \frac{1}{2}$ where the filament loses orientation \cite{ball2008orientable}.
With the  model in hand we address two key questions about the bacteriophage structure: (1)the osmotic pressure, and (2) the size of the isotropic region. 
 The calculation of the Cauchy stress tensor $\mathbb{T}$ associated with the total energy (\ref{total-energy}) follows a standard variational approach, formally expressed as $\mathbb{T}=\delta E/\delta \mathbf{x}$ are standard.
%  where one calculates 
%the energy variation $\delta E$ corresponding to the change in the deformation map $\mathbf{x}\longrightarrow \mathbf{x}+\delta\mathbf{x}$, for $\mathbf{x}\in\mathcal{B}$. These calculations, formally expressed as  
 The  pressure near the surface of the capsid is then given by $P=\boldsymbol{\nu}^T\mathbb{T}\boldsymbol{\nu}, $ with 
 $\boldsymbol{\nu}$ denoting the unit outer normal to the surface that on each tested conformation (TABLE I) has the form
\begin{align*}
%\sigma &= \frac{0.257}{L_pd}, \\
%\nu &= \ln\left(\frac{4}{\pi}c^2 - \frac{45}{8} + c^{-2}\right) - 1, \\
P(r) &= \frac{L_p}{\pi r^5}\cdot\frac{R_gT\rhoDNA}{\MDNA} \quad \text{with }r = 1,
\end{align*}
where $R_g$ is the universal gas constant; $T = 310$K is the temperature; $\rhoDNA$ is the density of (ds)DNA, and $\MDNA$ its  molar mass. (The ejection force at each point corresponds to the tangential component of $
\mathbb{T}\boldsymbol{\nu}$).
 The fifth column in Table II shows the predicted pressures. They fall within  the range of the experimentally measured forces on the capsid \cite{Evilevitch2003,Tzlil2003}, with bacteriophages $T4$ and $T5$ having smaller values. 
 
Next we estimated the size of the disordered (isotropic) core. In this study we assume that the capsid has the shape of a sphere equally truncated at the poles, with DNA spooled around its distinguished axis, with cylindrically arranged layering.  We prescribe the vector fields according to  the geometry,  so that   
 $\bn={\boldsymbol e}_\theta, \bm=-{\boldsymbol e}_r$ and $\bp={\boldsymbol e}_z, $ with layer locations  given by the  level surfaces of $\omega$ and $\gamma$ (\ref{level-surfaces}), and such that $\bm =q\nabla\omega$ and $\bp=q\nabla\theta$. 
  Substituting these fields into  (\ref{energy-na}), we minimize the resulting energy (\ref{total-energy}) with respect to the unknown isotropic core radius $r_0$. 
   TABLE \ref{results-table} shows the the data used in these calculations (column 2), the predicted value (column 3) and the estimated error(column 4). 
   %The values are consistent with experimental measurements for bacteriophage $P22$. 
 
%The viruses shown in the tables have been chosen because of  the small protein core size, except for $\epsilon 15$. %, which we have neglected in the calculation. 
 % (See Fig.\ref{EM_Illustrations}).
\begin{table}%[H] 
	\begin{center}
		\begin{tabular}{|c|c|c|c|c|c|c|c|}
			\hline
			Virus  & $L_p$ & $d$ & $c$&$ L$ (nm)& $R$ (nm)&$r_0/R$ \\ \hline\hline
			T4 & 1.32 & 0.06 & 21.37 & 55047.6&40.0&0.5500\\ \hline
			T5 & 1.39 & 0.07 & 17.85&39423.8&42.0&0.4286 \\ \hline
			T7 & 2.03 & 0.10 & 18.17&12932.0&26.05&0.5889 \\ \hline
			$\epsilon$15  & 1.9 & 0.09 & 13.98&12846.0&28.37&0.5735 \\ \hline
		\end{tabular}
	\caption{Physical measurements of four different bacteriophages.The symbol $L_P$ denotes the persistence length of a DNA chain of length $L$, effective diameter $d$, molar concentration $c$ in a sphere-like capsid of radius $R$ with a measured radius $r_0$ of the disordered core. T4 \cite{leiman2003structure,olson2001structure}; T5 \cite{Effantin2006}; T7 \cite{Cerritelli1997}; $\epsilon$15 \cite{Jiang2006}.
		%	\note{Add references and include units of $L_p$, $d$ and $c$}
		}
	\label{data}
	\end{center}	\label{DNA-data-table}
\end{table}
% The vector $\mathbf{r}_0$ gives the location of the filament tip at the entrance of the capsid.
Ou proposed model can be easily extended to incorporate other assumptions and data. First  the model is purely geometrical and mechanical, although it implicitly accounts for electrostatic repulsion and ionic effects by incorporating  the value of the effective diameter of the DNA under confinement. A model explicitly accounting for electrostatic and ionic effects, including those of the environment, and that may predict higher values of the pressure,   is currently being developed by the authors. Second as a consequence of the unit director length constraint, the curve $\mathbf{r}(s)$ must satisfy $|\mathbf r'(s)|=1,$ which states that the DNA filament is inextensible.  However, there is evidence that under certain conditions, DNA may stretch to the point of breaking down its double helix structure (reviewed in \cite{bustamante2003ten}). In future work, the constraint of unit director length will be relaxed by including the penalty term $(|\mathbf{n}(\mathbf{x})|^2-1)^2$.
Third, the model also allows for departure from the deterministic packing by the incorporation of appropriate random noise terms which will help explain the formation of knots in some bacteriophages \cite{Arsuaga2002b,Arsuaga2005}.  Fourth, the description of the capsid domain can also be modified to cases when the protein complex that includes the molecular motor protrudes inside the volume determined by the capsid. Fifth, our model is also amenable to be combined with parameter determining optimization methods, making it appropriate to broader designing features required in applications in medicine and biotechnology.  Sixth, chiral configurations can also be treated  extending the current model to include chiral effects in the energy and taking into account the imprinted protein twist configurations on the boundary of the capsid.  Finally, the present model neglects thermal effects that may be relevant to certain type of viruses (e.g. the human Herpes HSV-1) that will be addressed in future works. 
%\begin{center}
%\begin{table}[h]
%\begin{tabular}{|c|c|c|c|c|}\hline
%Virus & Measured Core Size & Predicted Core Size & Error & P(atm) \\ \hline
%T4 & 0.5500 & 0.5348 & 2.76\% & 44.10 \\ \hline
%T5 & 0.4286 & 0.4268 & 0.40\% & 46.42 \\ \hline
%T7 & 0.5889 & 0.5712 & 3.00\%& 67.72 \\ \hline
%$\epsilon 15$ & 0.5735 & 0.569 & 0.639\% & 62.18 \\ \hline
%\end{tabular}
\begin{center}
	\begin{table}[h]
		\begin{tabular}{|c|c|c|c|c|}\hline
			Virus & Measured Core Size & Predicted Core Size & Error & P(atm) \\ \hline
			T4 & 0.5500 & 0.5348 & 2.76\% & 28.70 \\ \hline
			T5 & 0.4286 & 0.4268 & 0.40\% & 30.17 \\ \hline
			T7 & 0.5889 & 0.5712 & 3.00\%& 44.02 \\ \hline
			$\epsilon 15$ & 0.5735 & 0.569 & 0.639\% & 40.41 \\ \hline
		\end{tabular}
\caption{Predicted size of the disordered phase and osmotic pressure for different bacteriophages. The table lists the measured and predicted radii of the isotropic for several viruses. The values are scaled by the corresponding capsid radii as listed in Table I, %(\ref{DNA-data-table}).
	 The percentage error is the difference between the measured and predicted size of the isotropic phase, divided by the measured core size. In the calculation for T5 and $\epsilon15$, the expressions of the material parameters $\nu$ \cite{onsager1949} and $\sigma$ \cite{DK84} correspond to the approximation in the low concentration regime. The evaluations for the other entries in the table are taken at the high concentration limits \cite{vanRoij2005-1,vanRoij2005-2,priezjev2000}. The last column lists the pressure calculated near the capsid boundary.}
\label{results-table}
\end{table}
\end{center}
%\end{widetext}
\bibliography{gel-PRL-draft}

\begin{thebibliography}{56}%
\makeatletter
\providecommand \@ifxundefined [1]{%
 \@ifx{#1\undefined}
}%
\providecommand \@ifnum [1]{%
 \ifnum #1\expandafter \@firstoftwo
 \else \expandafter \@secondoftwo
 \fi
}%
\providecommand \@ifx [1]{%
 \ifx #1\expandafter \@firstoftwo
 \else \expandafter \@secondoftwo
 \fi
}%
\providecommand \natexlab [1]{#1}%
\providecommand \enquote  [1]{``#1''}%
\providecommand \bibnamefont  [1]{#1}%
\providecommand \bibfnamefont [1]{#1}%
\providecommand \citenamefont [1]{#1}%
\providecommand \href@noop [0]{\@secondoftwo}%
\providecommand \href [0]{\begingroup \@sanitize@url \@href}%
\providecommand \@href[1]{\@@startlink{#1}\@@href}%
\providecommand \@@href[1]{\endgroup#1\@@endlink}%
\providecommand \@sanitize@url [0]{\catcode `\\12\catcode `\$12\catcode
  `\&12\catcode `\#12\catcode `\^12\catcode `\_12\catcode `\%12\relax}%
\providecommand \@@startlink[1]{}%
\providecommand \@@endlink[0]{}%
\providecommand \url  [0]{\begingroup\@sanitize@url \@url }%
\providecommand \@url [1]{\endgroup\@href {#1}{\urlprefix }}%
\providecommand \urlprefix  [0]{URL }%
\providecommand \Eprint [0]{\href }%
\providecommand \doibase [0]{http://dx.doi.org/}%
\providecommand \selectlanguage [0]{\@gobble}%
\providecommand \bibinfo  [0]{\@secondoftwo}%
\providecommand \bibfield  [0]{\@secondoftwo}%
\providecommand \translation [1]{[#1]}%
\providecommand \BibitemOpen [0]{}%
\providecommand \bibitemStop [0]{}%
\providecommand \bibitemNoStop [0]{.\EOS\space}%
\providecommand \EOS [0]{\spacefactor3000\relax}%
\providecommand \BibitemShut  [1]{\csname bibitem#1\endcsname}%
\let\auto@bib@innerbib\@empty
%</preamble>
\bibitem [{\citenamefont {Haq}\ \emph {et~al.}(2012)\citenamefont {Haq},
  \citenamefont {Chaudhry}, \citenamefont {Akhtar}, \citenamefont {Andleeb},\
  and\ \citenamefont {Qadri}}]{haq2012bacteriophages}%
  \BibitemOpen
  \bibfield  {author} {\bibinfo {author} {\bibfnamefont {I.~U.}\ \bibnamefont
  {Haq}}, \bibinfo {author} {\bibfnamefont {W.~N.}\ \bibnamefont {Chaudhry}},
  \bibinfo {author} {\bibfnamefont {M.~N.}\ \bibnamefont {Akhtar}}, \bibinfo
  {author} {\bibfnamefont {S.}~\bibnamefont {Andleeb}}, \ and\ \bibinfo
  {author} {\bibfnamefont {I.}~\bibnamefont {Qadri}},\ }\href@noop {}
  {\bibfield  {journal} {\bibinfo  {journal} {Virology journal}\ }\textbf
  {\bibinfo {volume} {9}},\ \bibinfo {pages} {9} (\bibinfo {year}
  {2012})}\BibitemShut {NoStop}%
\bibitem [{\citenamefont {Sulakvelidze}\ \emph {et~al.}(2001)\citenamefont
  {Sulakvelidze}, \citenamefont {Alavidze},\ and\ \citenamefont
  {Morris}}]{Sulakvelidze2001}%
  \BibitemOpen
  \bibfield  {author} {\bibinfo {author} {\bibfnamefont {A.}~\bibnamefont
  {Sulakvelidze}}, \bibinfo {author} {\bibfnamefont {Z.}~\bibnamefont
  {Alavidze}}, \ and\ \bibinfo {author} {\bibfnamefont {J.~G.}\ \bibnamefont
  {Morris}},\ }\href@noop {} {\bibfield  {journal} {\bibinfo  {journal}
  {Antimicrobial agents and chemotherapy}\ }\textbf {\bibinfo {volume} {45}},\
  \bibinfo {pages} {649} (\bibinfo {year} {2001})}\BibitemShut {NoStop}%
\bibitem [{\citenamefont {Smith}\ \emph {et~al.}(2001)\citenamefont {Smith},
  \citenamefont {Tans}, \citenamefont {Smith}, \citenamefont {Grimes},
  \citenamefont {Anderson},\ and\ \citenamefont {Bustamante}}]{Smith2001}%
  \BibitemOpen
  \bibfield  {author} {\bibinfo {author} {\bibfnamefont {D.~E.}\ \bibnamefont
  {Smith}}, \bibinfo {author} {\bibfnamefont {S.~J.}\ \bibnamefont {Tans}},
  \bibinfo {author} {\bibfnamefont {S.~B.}\ \bibnamefont {Smith}}, \bibinfo
  {author} {\bibfnamefont {S.}~\bibnamefont {Grimes}}, \bibinfo {author}
  {\bibfnamefont {D.~L.}\ \bibnamefont {Anderson}}, \ and\ \bibinfo {author}
  {\bibfnamefont {C.}~\bibnamefont {Bustamante}},\ }\href@noop {} {\bibfield
  {journal} {\bibinfo  {journal} {Nature}\ }\textbf {\bibinfo {volume} {413}},\
  \bibinfo {pages} {748} (\bibinfo {year} {2001})}\BibitemShut {NoStop}%
\bibitem [{\citenamefont {Leforestier}\ and\ \citenamefont
  {Livolant}(2010)}]{Leforestier2010}%
  \BibitemOpen
  \bibfield  {author} {\bibinfo {author} {\bibfnamefont {A.}~\bibnamefont
  {Leforestier}}\ and\ \bibinfo {author} {\bibfnamefont {F.}~\bibnamefont
  {Livolant}},\ }\href@noop {} {\bibfield  {journal} {\bibinfo  {journal}
  {Journal of molecular biology}\ }\textbf {\bibinfo {volume} {396}},\ \bibinfo
  {pages} {384} (\bibinfo {year} {2010})}\BibitemShut {NoStop}%
\bibitem [{\citenamefont {Liu}\ \emph {et~al.}(2014)\citenamefont {Liu},
  \citenamefont {Sae-Ueng}, \citenamefont {Li}, \citenamefont {Lander},
  \citenamefont {Zuo}, \citenamefont {J{\"o}nsson}, \citenamefont {Rau},
  \citenamefont {Shefer},\ and\ \citenamefont {Evilevitch}}]{Liu2014solid}%
  \BibitemOpen
  \bibfield  {author} {\bibinfo {author} {\bibfnamefont {T.}~\bibnamefont
  {Liu}}, \bibinfo {author} {\bibfnamefont {U.}~\bibnamefont {Sae-Ueng}},
  \bibinfo {author} {\bibfnamefont {D.}~\bibnamefont {Li}}, \bibinfo {author}
  {\bibfnamefont {G.~C.}\ \bibnamefont {Lander}}, \bibinfo {author}
  {\bibfnamefont {X.}~\bibnamefont {Zuo}}, \bibinfo {author} {\bibfnamefont
  {B.}~\bibnamefont {J{\"o}nsson}}, \bibinfo {author} {\bibfnamefont
  {D.}~\bibnamefont {Rau}}, \bibinfo {author} {\bibfnamefont {I.}~\bibnamefont
  {Shefer}}, \ and\ \bibinfo {author} {\bibfnamefont {A.}~\bibnamefont
  {Evilevitch}},\ }\href@noop {} {\bibfield  {journal} {\bibinfo  {journal}
  {Proceedings of the National Academy of Sciences}\ }\textbf {\bibinfo
  {volume} {111}},\ \bibinfo {pages} {14675} (\bibinfo {year}
  {2014})}\BibitemShut {NoStop}%
\bibitem [{\citenamefont {Sae-Ueng}\ \emph {et~al.}(2014)\citenamefont
  {Sae-Ueng}, \citenamefont {Li}, \citenamefont {Zuo}, \citenamefont {Huffman},
  \citenamefont {Homa}, \citenamefont {Rau},\ and\ \citenamefont
  {Evilevitch}}]{Sae2014}%
  \BibitemOpen
  \bibfield  {author} {\bibinfo {author} {\bibfnamefont {U.}~\bibnamefont
  {Sae-Ueng}}, \bibinfo {author} {\bibfnamefont {D.}~\bibnamefont {Li}},
  \bibinfo {author} {\bibfnamefont {X.}~\bibnamefont {Zuo}}, \bibinfo {author}
  {\bibfnamefont {J.~B.}\ \bibnamefont {Huffman}}, \bibinfo {author}
  {\bibfnamefont {F.~L.}\ \bibnamefont {Homa}}, \bibinfo {author}
  {\bibfnamefont {D.}~\bibnamefont {Rau}}, \ and\ \bibinfo {author}
  {\bibfnamefont {A.}~\bibnamefont {Evilevitch}},\ }\href@noop {} {\bibfield
  {journal} {\bibinfo  {journal} {Nature chemical biology}\ }\textbf {\bibinfo
  {volume} {10}},\ \bibinfo {pages} {861} (\bibinfo {year} {2014})}\BibitemShut
  {NoStop}%
\bibitem [{\citenamefont {Kellenberger}\ \emph
  {et~al.}(1986{\natexlab{a}})\citenamefont {Kellenberger}, \citenamefont
  {Carlemalm}, \citenamefont {Sechaud}, \citenamefont {Ryter},\ and\
  \citenamefont {De~Haller}}]{Kellenberger1986}%
  \BibitemOpen
  \bibfield  {author} {\bibinfo {author} {\bibfnamefont {E.}~\bibnamefont
  {Kellenberger}}, \bibinfo {author} {\bibfnamefont {E.}~\bibnamefont
  {Carlemalm}}, \bibinfo {author} {\bibfnamefont {J.}~\bibnamefont {Sechaud}},
  \bibinfo {author} {\bibfnamefont {A.}~\bibnamefont {Ryter}}, \ and\ \bibinfo
  {author} {\bibfnamefont {G.}~\bibnamefont {De~Haller}},\ }in\ \href@noop {}
  {\emph {\bibinfo {booktitle} {Bacterial chromatin}}}\ (\bibinfo  {publisher}
  {Springer},\ \bibinfo {year} {1986})\ pp.\ \bibinfo {pages}
  {11--25}\BibitemShut {NoStop}%
\bibitem [{\citenamefont {Evilevitch}\ \emph {et~al.}(2003)\citenamefont
  {Evilevitch}, \citenamefont {Lavelle}, \citenamefont {Knobler}, \citenamefont
  {Raspaud},\ and\ \citenamefont {Gelbart}}]{Evilevitch2003}%
  \BibitemOpen
  \bibfield  {author} {\bibinfo {author} {\bibfnamefont {A.}~\bibnamefont
  {Evilevitch}}, \bibinfo {author} {\bibfnamefont {L.}~\bibnamefont {Lavelle}},
  \bibinfo {author} {\bibfnamefont {C.~M.}\ \bibnamefont {Knobler}}, \bibinfo
  {author} {\bibfnamefont {E.}~\bibnamefont {Raspaud}}, \ and\ \bibinfo
  {author} {\bibfnamefont {W.~M.}\ \bibnamefont {Gelbart}},\ }\href@noop {}
  {\bibfield  {journal} {\bibinfo  {journal} {Proceedings of the National
  Academy of Sciences}\ }\textbf {\bibinfo {volume} {100}},\ \bibinfo {pages}
  {9292} (\bibinfo {year} {2003})}\BibitemShut {NoStop}%
\bibitem [{\citenamefont {Jeembaeva}\ \emph {et~al.}(2008)\citenamefont
  {Jeembaeva}, \citenamefont {Castelnovo}, \citenamefont {Larsson},\ and\
  \citenamefont {Evilevitch}}]{Jeembaeva2008}%
  \BibitemOpen
  \bibfield  {author} {\bibinfo {author} {\bibfnamefont {M.}~\bibnamefont
  {Jeembaeva}}, \bibinfo {author} {\bibfnamefont {M.}~\bibnamefont
  {Castelnovo}}, \bibinfo {author} {\bibfnamefont {F.}~\bibnamefont {Larsson}},
  \ and\ \bibinfo {author} {\bibfnamefont {A.}~\bibnamefont {Evilevitch}},\
  }\href@noop {} {\bibfield  {journal} {\bibinfo  {journal} {Journal of
  molecular biology}\ }\textbf {\bibinfo {volume} {381}},\ \bibinfo {pages}
  {310} (\bibinfo {year} {2008})}\BibitemShut {NoStop}%
\bibitem [{\citenamefont {Riemer}\ and\ \citenamefont
  {Bloomfield}(1978)}]{RiemerBloomfield1978}%
  \BibitemOpen
  \bibfield  {author} {\bibinfo {author} {\bibfnamefont {S.~C.}\ \bibnamefont
  {Riemer}}\ and\ \bibinfo {author} {\bibfnamefont {V.~A.}\ \bibnamefont
  {Bloomfield}},\ }\href@noop {} {\bibfield  {journal} {\bibinfo  {journal}
  {Biopolymers}\ }\textbf {\bibinfo {volume} {17}},\ \bibinfo {pages} {785}
  (\bibinfo {year} {1978})}\BibitemShut {NoStop}%
\bibitem [{\citenamefont {Leforestier}\ and\ \citenamefont
  {Livolant}(2009)}]{Leforestier2009}%
  \BibitemOpen
  \bibfield  {author} {\bibinfo {author} {\bibfnamefont {A.}~\bibnamefont
  {Leforestier}}\ and\ \bibinfo {author} {\bibfnamefont {F.}~\bibnamefont
  {Livolant}},\ }\href@noop {} {\bibfield  {journal} {\bibinfo  {journal}
  {Proceedings of the National Academy of Sciences}\ }\textbf {\bibinfo
  {volume} {106}},\ \bibinfo {pages} {9157} (\bibinfo {year}
  {2009})}\BibitemShut {NoStop}%
\bibitem [{\citenamefont {Lepault}\ \emph {et~al.}(1987)\citenamefont
  {Lepault}, \citenamefont {Dubochet}, \citenamefont {Baschong},\ and\
  \citenamefont {Kellenberger}}]{Lepault1987}%
  \BibitemOpen
  \bibfield  {author} {\bibinfo {author} {\bibfnamefont {J.}~\bibnamefont
  {Lepault}}, \bibinfo {author} {\bibfnamefont {J.}~\bibnamefont {Dubochet}},
  \bibinfo {author} {\bibfnamefont {W.}~\bibnamefont {Baschong}}, \ and\
  \bibinfo {author} {\bibfnamefont {E.}~\bibnamefont {Kellenberger}},\
  }\href@noop {} {\bibfield  {journal} {\bibinfo  {journal} {The EMBO journal}\
  }\textbf {\bibinfo {volume} {6}},\ \bibinfo {pages} {1507} (\bibinfo {year}
  {1987})}\BibitemShut {NoStop}%
\bibitem [{\citenamefont {Reith}\ \emph {et~al.}(2012)\citenamefont {Reith},
  \citenamefont {Cifra}, \citenamefont {Stasiak},\ and\ \citenamefont
  {Virnau}}]{Reith2012}%
  \BibitemOpen
  \bibfield  {author} {\bibinfo {author} {\bibfnamefont {D.}~\bibnamefont
  {Reith}}, \bibinfo {author} {\bibfnamefont {P.}~\bibnamefont {Cifra}},
  \bibinfo {author} {\bibfnamefont {A.}~\bibnamefont {Stasiak}}, \ and\
  \bibinfo {author} {\bibfnamefont {P.}~\bibnamefont {Virnau}},\ }\href@noop {}
  {\bibfield  {journal} {\bibinfo  {journal} {Nucleic acids research}\ }\textbf
  {\bibinfo {volume} {40}},\ \bibinfo {pages} {5129} (\bibinfo {year}
  {2012})}\BibitemShut {NoStop}%
\bibitem [{\citenamefont {Rill}(1986)}]{Rill1986}%
  \BibitemOpen
  \bibfield  {author} {\bibinfo {author} {\bibfnamefont {R.~L.}\ \bibnamefont
  {Rill}},\ }\href@noop {} {\bibfield  {journal} {\bibinfo  {journal}
  {Proceedings of the National Academy of Sciences}\ }\textbf {\bibinfo
  {volume} {83}},\ \bibinfo {pages} {342} (\bibinfo {year} {1986})}\BibitemShut
  {NoStop}%
\bibitem [{\citenamefont {Strzelecka}\ \emph {et~al.}(1988)\citenamefont
  {Strzelecka}, \citenamefont {Davidson},\ and\ \citenamefont
  {Rill}}]{Strzelecka1988}%
  \BibitemOpen
  \bibfield  {author} {\bibinfo {author} {\bibfnamefont {T.~E.}\ \bibnamefont
  {Strzelecka}}, \bibinfo {author} {\bibfnamefont {M.~W.}\ \bibnamefont
  {Davidson}}, \ and\ \bibinfo {author} {\bibfnamefont {R.~L.}\ \bibnamefont
  {Rill}},\ }\href@noop {} {\bibfield  {journal} {\bibinfo  {journal} {Nature}\
  }\textbf {\bibinfo {volume} {331}},\ \bibinfo {pages} {457} (\bibinfo {year}
  {1988})}\BibitemShut {NoStop}%
\bibitem [{\citenamefont {Park}\ \emph {et~al.}(2008)\citenamefont {Park},
  \citenamefont {Kang}, \citenamefont {Tortora}, \citenamefont {Nastishin},
  \citenamefont {Finotello}, \citenamefont {Kumar},\ and\ \citenamefont
  {Lavrentovich}}]{park2008self}%
  \BibitemOpen
  \bibfield  {author} {\bibinfo {author} {\bibfnamefont {H.-S.}\ \bibnamefont
  {Park}}, \bibinfo {author} {\bibfnamefont {S.-W.}\ \bibnamefont {Kang}},
  \bibinfo {author} {\bibfnamefont {L.}~\bibnamefont {Tortora}}, \bibinfo
  {author} {\bibfnamefont {Y.}~\bibnamefont {Nastishin}}, \bibinfo {author}
  {\bibfnamefont {D.}~\bibnamefont {Finotello}}, \bibinfo {author}
  {\bibfnamefont {S.}~\bibnamefont {Kumar}}, \ and\ \bibinfo {author}
  {\bibfnamefont {O.~D.}\ \bibnamefont {Lavrentovich}},\ }\href@noop {}
  {\bibfield  {journal} {\bibinfo  {journal} {The Journal of Physical Chemistry
  B}\ }\textbf {\bibinfo {volume} {112}},\ \bibinfo {pages} {16307} (\bibinfo
  {year} {2008})}\BibitemShut {NoStop}%
\bibitem [{\citenamefont {Livolant}(1991)}]{Livolant1991}%
  \BibitemOpen
  \bibfield  {author} {\bibinfo {author} {\bibfnamefont {F.}~\bibnamefont
  {Livolant}},\ }\href@noop {} {\bibfield  {journal} {\bibinfo  {journal}
  {Physica A: Statistical Mechanics and its Applications}\ }\textbf {\bibinfo
  {volume} {176}},\ \bibinfo {pages} {117} (\bibinfo {year}
  {1991})}\BibitemShut {NoStop}%
\bibitem [{\citenamefont {Leforestier}\ and\ \citenamefont
  {Livolant}(1993)}]{Leforestier1993}%
  \BibitemOpen
  \bibfield  {author} {\bibinfo {author} {\bibfnamefont {A.}~\bibnamefont
  {Leforestier}}\ and\ \bibinfo {author} {\bibfnamefont {F.}~\bibnamefont
  {Livolant}},\ }\href@noop {} {\bibfield  {journal} {\bibinfo  {journal}
  {Biophysical journal}\ }\textbf {\bibinfo {volume} {65}},\ \bibinfo {pages}
  {56} (\bibinfo {year} {1993})}\BibitemShut {NoStop}%
\bibitem [{\citenamefont {Leforestier}\ \emph {et~al.}(2008)\citenamefont
  {Leforestier}, \citenamefont {Brasiles}, \citenamefont {De~Frutos},
  \citenamefont {Raspaud}, \citenamefont {Letellier}, \citenamefont {Tavares},\
  and\ \citenamefont {Livolant}}]{leforestier2008bacteriophage}%
  \BibitemOpen
  \bibfield  {author} {\bibinfo {author} {\bibfnamefont {A.}~\bibnamefont
  {Leforestier}}, \bibinfo {author} {\bibfnamefont {S.}~\bibnamefont
  {Brasiles}}, \bibinfo {author} {\bibfnamefont {M.}~\bibnamefont {De~Frutos}},
  \bibinfo {author} {\bibfnamefont {E.}~\bibnamefont {Raspaud}}, \bibinfo
  {author} {\bibfnamefont {L.}~\bibnamefont {Letellier}}, \bibinfo {author}
  {\bibfnamefont {P.}~\bibnamefont {Tavares}}, \ and\ \bibinfo {author}
  {\bibfnamefont {F.}~\bibnamefont {Livolant}},\ }\href@noop {} {\bibfield
  {journal} {\bibinfo  {journal} {Journal of molecular biology}\ }\textbf
  {\bibinfo {volume} {384}},\ \bibinfo {pages} {730} (\bibinfo {year}
  {2008})}\BibitemShut {NoStop}%
\bibitem [{\citenamefont {Marenduzzo}\ \emph {et~al.}(2009)\citenamefont
  {Marenduzzo}, \citenamefont {Orlandini}, \citenamefont {Stasiak},
  \citenamefont {Tubiana}, \citenamefont {Micheletti} \emph
  {et~al.}}]{Marenduzzo2009}%
  \BibitemOpen
  \bibfield  {author} {\bibinfo {author} {\bibfnamefont {D.}~\bibnamefont
  {Marenduzzo}}, \bibinfo {author} {\bibfnamefont {E.}~\bibnamefont
  {Orlandini}}, \bibinfo {author} {\bibfnamefont {A.}~\bibnamefont {Stasiak}},
  \bibinfo {author} {\bibfnamefont {L.}~\bibnamefont {Tubiana}}, \bibinfo
  {author} {\bibfnamefont {C.}~\bibnamefont {Micheletti}},  \emph {et~al.},\
  }\href@noop {} {\bibfield  {journal} {\bibinfo  {journal} {Proceedings of the
  National Academy of Sciences}\ }\textbf {\bibinfo {volume} {106}},\ \bibinfo
  {pages} {22269} (\bibinfo {year} {2009})}\BibitemShut {NoStop}%
\bibitem [{\citenamefont {Cerritelli}\ \emph {et~al.}(1997)\citenamefont
  {Cerritelli}, \citenamefont {Cheng}, \citenamefont {Rosenberg}, \citenamefont
  {McPherson}, \citenamefont {Booy},\ and\ \citenamefont
  {Steven}}]{Cerritelli1997}%
  \BibitemOpen
  \bibfield  {author} {\bibinfo {author} {\bibfnamefont {M.~E.}\ \bibnamefont
  {Cerritelli}}, \bibinfo {author} {\bibfnamefont {N.}~\bibnamefont {Cheng}},
  \bibinfo {author} {\bibfnamefont {A.~H.}\ \bibnamefont {Rosenberg}}, \bibinfo
  {author} {\bibfnamefont {C.~E.}\ \bibnamefont {McPherson}}, \bibinfo {author}
  {\bibfnamefont {F.~P.}\ \bibnamefont {Booy}}, \ and\ \bibinfo {author}
  {\bibfnamefont {A.~C.}\ \bibnamefont {Steven}},\ }\href@noop {} {\bibfield
  {journal} {\bibinfo  {journal} {Cell}\ }\textbf {\bibinfo {volume} {91}},\
  \bibinfo {pages} {271} (\bibinfo {year} {1997})}\BibitemShut {NoStop}%
\bibitem [{\citenamefont {Earnshaw}\ and\ \citenamefont
  {Casjens}(1980)}]{Earnshaw1980}%
  \BibitemOpen
  \bibfield  {author} {\bibinfo {author} {\bibfnamefont {W.~C.}\ \bibnamefont
  {Earnshaw}}\ and\ \bibinfo {author} {\bibfnamefont {S.~R.}\ \bibnamefont
  {Casjens}},\ }\href@noop {} {\bibfield  {journal} {\bibinfo  {journal}
  {Cell}\ }\textbf {\bibinfo {volume} {21}},\ \bibinfo {pages} {319} (\bibinfo
  {year} {1980})}\BibitemShut {NoStop}%
\bibitem [{\citenamefont {Hud}(1995)}]{Hud1995}%
  \BibitemOpen
  \bibfield  {author} {\bibinfo {author} {\bibfnamefont {N.~V.}\ \bibnamefont
  {Hud}},\ }\href@noop {} {\bibfield  {journal} {\bibinfo  {journal}
  {Biophysical journal}\ }\textbf {\bibinfo {volume} {69}},\ \bibinfo {pages}
  {1355} (\bibinfo {year} {1995})}\BibitemShut {NoStop}%
\bibitem [{\citenamefont {Petrov}\ \emph {et~al.}(2007)\citenamefont {Petrov},
  \citenamefont {Boz},\ and\ \citenamefont {Harvey}}]{Petrov2007a}%
  \BibitemOpen
  \bibfield  {author} {\bibinfo {author} {\bibfnamefont {A.~S.}\ \bibnamefont
  {Petrov}}, \bibinfo {author} {\bibfnamefont {M.~B.}\ \bibnamefont {Boz}}, \
  and\ \bibinfo {author} {\bibfnamefont {S.~C.}\ \bibnamefont {Harvey}},\
  }\href@noop {} {\bibfield  {journal} {\bibinfo  {journal} {Journal of
  structural biology}\ }\textbf {\bibinfo {volume} {160}},\ \bibinfo {pages}
  {241} (\bibinfo {year} {2007})}\BibitemShut {NoStop}%
\bibitem [{\citenamefont {Serwer}\ \emph {et~al.}(1992)\citenamefont {Serwer},
  \citenamefont {Hayes},\ and\ \citenamefont {Watson}}]{Serwer1992}%
  \BibitemOpen
  \bibfield  {author} {\bibinfo {author} {\bibfnamefont {P.}~\bibnamefont
  {Serwer}}, \bibinfo {author} {\bibfnamefont {S.~J.}\ \bibnamefont {Hayes}}, \
  and\ \bibinfo {author} {\bibfnamefont {R.~H.}\ \bibnamefont {Watson}},\
  }\href@noop {} {\bibfield  {journal} {\bibinfo  {journal} {Journal of
  molecular biology}\ }\textbf {\bibinfo {volume} {223}},\ \bibinfo {pages}
  {999} (\bibinfo {year} {1992})}\BibitemShut {NoStop}%
\bibitem [{\citenamefont {Arsuaga}\ and\ \citenamefont
  {Diao}(2008)}]{Arsuaga2008}%
  \BibitemOpen
  \bibfield  {author} {\bibinfo {author} {\bibfnamefont {J.}~\bibnamefont
  {Arsuaga}}\ and\ \bibinfo {author} {\bibfnamefont {Y.}~\bibnamefont {Diao}},\
  }\href@noop {} {\bibfield  {journal} {\bibinfo  {journal} {Computational and
  Mathematical Methods in Medicine}\ }\textbf {\bibinfo {volume} {9}},\
  \bibinfo {pages} {303} (\bibinfo {year} {2008})}\BibitemShut {NoStop}%
\bibitem [{\citenamefont {Arsuaga}\ \emph
  {et~al.}(2002{\natexlab{a}})\citenamefont {Arsuaga}, \citenamefont {Tan},
  \citenamefont {Vazquez}, \citenamefont {Harvey} \emph
  {et~al.}}]{Arsuaga2002a}%
  \BibitemOpen
  \bibfield  {author} {\bibinfo {author} {\bibfnamefont {J.}~\bibnamefont
  {Arsuaga}}, \bibinfo {author} {\bibfnamefont {R.~K.-Z.}\ \bibnamefont {Tan}},
  \bibinfo {author} {\bibfnamefont {M.}~\bibnamefont {Vazquez}}, \bibinfo
  {author} {\bibfnamefont {S.~C.}\ \bibnamefont {Harvey}},  \emph {et~al.},\
  }\href@noop {} {\bibfield  {journal} {\bibinfo  {journal} {Biophysical
  chemistry}\ }\textbf {\bibinfo {volume} {101}},\ \bibinfo {pages} {475}
  (\bibinfo {year} {2002}{\natexlab{a}})}\BibitemShut {NoStop}%
\bibitem [{\citenamefont {Arsuaga}\ \emph
  {et~al.}(2002{\natexlab{b}})\citenamefont {Arsuaga}, \citenamefont
  {V{\'a}zquez}, \citenamefont {Trigueros}, \citenamefont {Roca} \emph
  {et~al.}}]{Arsuaga2002b}%
  \BibitemOpen
  \bibfield  {author} {\bibinfo {author} {\bibfnamefont {J.}~\bibnamefont
  {Arsuaga}}, \bibinfo {author} {\bibfnamefont {M.}~\bibnamefont
  {V{\'a}zquez}}, \bibinfo {author} {\bibfnamefont {S.}~\bibnamefont
  {Trigueros}}, \bibinfo {author} {\bibfnamefont {J.}~\bibnamefont {Roca}},
  \emph {et~al.},\ }\href@noop {} {\bibfield  {journal} {\bibinfo  {journal}
  {Proceedings of the National Academy of Sciences}\ }\textbf {\bibinfo
  {volume} {99}},\ \bibinfo {pages} {5373} (\bibinfo {year}
  {2002}{\natexlab{b}})}\BibitemShut {NoStop}%
\bibitem [{\citenamefont {Arsuaga}\ \emph {et~al.}(2005)\citenamefont
  {Arsuaga}, \citenamefont {Vazquez}, \citenamefont {McGuirk}, \citenamefont
  {Trigueros}, \citenamefont {Roca} \emph {et~al.}}]{Arsuaga2005}%
  \BibitemOpen
  \bibfield  {author} {\bibinfo {author} {\bibfnamefont {J.}~\bibnamefont
  {Arsuaga}}, \bibinfo {author} {\bibfnamefont {M.}~\bibnamefont {Vazquez}},
  \bibinfo {author} {\bibfnamefont {P.}~\bibnamefont {McGuirk}}, \bibinfo
  {author} {\bibfnamefont {S.}~\bibnamefont {Trigueros}}, \bibinfo {author}
  {\bibfnamefont {J.}~\bibnamefont {Roca}},  \emph {et~al.},\ }\href@noop {}
  {\bibfield  {journal} {\bibinfo  {journal} {Proceedings of the National
  Academy of Sciences of the United States of America}\ }\textbf {\bibinfo
  {volume} {102}},\ \bibinfo {pages} {9165} (\bibinfo {year}
  {2005})}\BibitemShut {NoStop}%
\bibitem [{\citenamefont {Marenduzzo}\ \emph {et~al.}(2013)\citenamefont
  {Marenduzzo}, \citenamefont {Micheletti}, \citenamefont {Orlandini} \emph
  {et~al.}}]{Marenduzzo2013}%
  \BibitemOpen
  \bibfield  {author} {\bibinfo {author} {\bibfnamefont {D.}~\bibnamefont
  {Marenduzzo}}, \bibinfo {author} {\bibfnamefont {C.}~\bibnamefont
  {Micheletti}}, \bibinfo {author} {\bibfnamefont {E.}~\bibnamefont
  {Orlandini}},  \emph {et~al.},\ }\href@noop {} {\bibfield  {journal}
  {\bibinfo  {journal} {Proceedings of the National Academy of Sciences}\
  }\textbf {\bibinfo {volume} {110}},\ \bibinfo {pages} {20081} (\bibinfo
  {year} {2013})}\BibitemShut {NoStop}%
\bibitem [{\citenamefont {Rollins}\ \emph {et~al.}(2008)\citenamefont
  {Rollins}, \citenamefont {Petrov},\ and\ \citenamefont
  {Harvey}}]{Rollins2008}%
  \BibitemOpen
  \bibfield  {author} {\bibinfo {author} {\bibfnamefont {G.~C.}\ \bibnamefont
  {Rollins}}, \bibinfo {author} {\bibfnamefont {A.~S.}\ \bibnamefont {Petrov}},
  \ and\ \bibinfo {author} {\bibfnamefont {S.~C.}\ \bibnamefont {Harvey}},\
  }\href@noop {} {\bibfield  {journal} {\bibinfo  {journal} {Biophysical
  journal}\ }\textbf {\bibinfo {volume} {94}},\ \bibinfo {pages} {L38}
  (\bibinfo {year} {2008})}\BibitemShut {NoStop}%
\bibitem [{\citenamefont {Spakowitz}\ and\ \citenamefont
  {Wang}(2005)}]{Spakowitz2005}%
  \BibitemOpen
  \bibfield  {author} {\bibinfo {author} {\bibfnamefont {A.~J.}\ \bibnamefont
  {Spakowitz}}\ and\ \bibinfo {author} {\bibfnamefont {Z.-G.}\ \bibnamefont
  {Wang}},\ }\href@noop {} {\bibfield  {journal} {\bibinfo  {journal}
  {Biophysical journal}\ }\textbf {\bibinfo {volume} {88}},\ \bibinfo {pages}
  {3912} (\bibinfo {year} {2005})}\BibitemShut {NoStop}%
\bibitem [{\citenamefont {Kellenberger}\ \emph
  {et~al.}(1986{\natexlab{b}})\citenamefont {Kellenberger}, \citenamefont
  {Carlemalm}, \citenamefont {Sechaud}, \citenamefont {Ryter},\ and\
  \citenamefont {De~Haller}}]{kellenberger1986considerations}%
  \BibitemOpen
  \bibfield  {author} {\bibinfo {author} {\bibfnamefont {E.}~\bibnamefont
  {Kellenberger}}, \bibinfo {author} {\bibfnamefont {E.}~\bibnamefont
  {Carlemalm}}, \bibinfo {author} {\bibfnamefont {J.}~\bibnamefont {Sechaud}},
  \bibinfo {author} {\bibfnamefont {A.}~\bibnamefont {Ryter}}, \ and\ \bibinfo
  {author} {\bibfnamefont {G.}~\bibnamefont {De~Haller}},\ }\href@noop {}
  {\bibfield  {journal} {\bibinfo  {journal} {Bacterial chromatin}\ }\textbf
  {\bibinfo {volume} {1}},\ \bibinfo {pages} {11} (\bibinfo {year}
  {1986}{\natexlab{b}})}\BibitemShut {NoStop}%
\bibitem [{\citenamefont {Comolli}\ \emph {et~al.}(2008)\citenamefont
  {Comolli}, \citenamefont {Spakowitz}, \citenamefont {Siegerist},
  \citenamefont {Jardine}, \citenamefont {Grimes}, \citenamefont {Anderson},
  \citenamefont {Bustamante},\ and\ \citenamefont {Downing}}]{Comolli2008}%
  \BibitemOpen
  \bibfield  {author} {\bibinfo {author} {\bibfnamefont {L.~R.}\ \bibnamefont
  {Comolli}}, \bibinfo {author} {\bibfnamefont {A.~J.}\ \bibnamefont
  {Spakowitz}}, \bibinfo {author} {\bibfnamefont {C.~E.}\ \bibnamefont
  {Siegerist}}, \bibinfo {author} {\bibfnamefont {P.~J.}\ \bibnamefont
  {Jardine}}, \bibinfo {author} {\bibfnamefont {S.}~\bibnamefont {Grimes}},
  \bibinfo {author} {\bibfnamefont {D.~L.}\ \bibnamefont {Anderson}}, \bibinfo
  {author} {\bibfnamefont {C.}~\bibnamefont {Bustamante}}, \ and\ \bibinfo
  {author} {\bibfnamefont {K.~H.}\ \bibnamefont {Downing}},\ }\href@noop {}
  {\bibfield  {journal} {\bibinfo  {journal} {Virology}\ }\textbf {\bibinfo
  {volume} {371}},\ \bibinfo {pages} {267} (\bibinfo {year}
  {2008})}\BibitemShut {NoStop}%
\bibitem [{\citenamefont {Chang}\ \emph {et~al.}(2006)\citenamefont {Chang},
  \citenamefont {Weigele}, \citenamefont {King}, \citenamefont {Chiu},\ and\
  \citenamefont {Jiang}}]{Chang2006}%
  \BibitemOpen
  \bibfield  {author} {\bibinfo {author} {\bibfnamefont {J.}~\bibnamefont
  {Chang}}, \bibinfo {author} {\bibfnamefont {P.}~\bibnamefont {Weigele}},
  \bibinfo {author} {\bibfnamefont {J.}~\bibnamefont {King}}, \bibinfo {author}
  {\bibfnamefont {W.}~\bibnamefont {Chiu}}, \ and\ \bibinfo {author}
  {\bibfnamefont {W.}~\bibnamefont {Jiang}},\ }\href@noop {} {\bibfield
  {journal} {\bibinfo  {journal} {Structure}\ }\textbf {\bibinfo {volume}
  {14}},\ \bibinfo {pages} {1073} (\bibinfo {year} {2006})}\BibitemShut
  {NoStop}%
\bibitem [{\citenamefont {Effantin}\ \emph {et~al.}(2006)\citenamefont
  {Effantin}, \citenamefont {Boulanger}, \citenamefont {Neumann}, \citenamefont
  {Letellier},\ and\ \citenamefont {Conway}}]{Effantin2006}%
  \BibitemOpen
  \bibfield  {author} {\bibinfo {author} {\bibfnamefont {G.}~\bibnamefont
  {Effantin}}, \bibinfo {author} {\bibfnamefont {P.}~\bibnamefont {Boulanger}},
  \bibinfo {author} {\bibfnamefont {E.}~\bibnamefont {Neumann}}, \bibinfo
  {author} {\bibfnamefont {L.}~\bibnamefont {Letellier}}, \ and\ \bibinfo
  {author} {\bibfnamefont {J.}~\bibnamefont {Conway}},\ }\href@noop {}
  {\bibfield  {journal} {\bibinfo  {journal} {Journal of molecular biology}\
  }\textbf {\bibinfo {volume} {361}},\ \bibinfo {pages} {993} (\bibinfo {year}
  {2006})}\BibitemShut {NoStop}%
\bibitem [{\citenamefont {Jiang}\ \emph {et~al.}(2006)\citenamefont {Jiang},
  \citenamefont {Chang}, \citenamefont {Jakana}, \citenamefont {Weigele},
  \citenamefont {King},\ and\ \citenamefont {Chiu}}]{Jiang2006}%
  \BibitemOpen
  \bibfield  {author} {\bibinfo {author} {\bibfnamefont {W.}~\bibnamefont
  {Jiang}}, \bibinfo {author} {\bibfnamefont {J.}~\bibnamefont {Chang}},
  \bibinfo {author} {\bibfnamefont {J.}~\bibnamefont {Jakana}}, \bibinfo
  {author} {\bibfnamefont {P.}~\bibnamefont {Weigele}}, \bibinfo {author}
  {\bibfnamefont {J.}~\bibnamefont {King}}, \ and\ \bibinfo {author}
  {\bibfnamefont {W.}~\bibnamefont {Chiu}},\ }\href@noop {} {\bibfield
  {journal} {\bibinfo  {journal} {Nature}\ }\textbf {\bibinfo {volume} {439}},\
  \bibinfo {pages} {612} (\bibinfo {year} {2006})}\BibitemShut {NoStop}%
\bibitem [{\citenamefont {Molineux}\ and\ \citenamefont
  {Panja}(2013)}]{Molineux2013}%
  \BibitemOpen
  \bibfield  {author} {\bibinfo {author} {\bibfnamefont {I.~J.}\ \bibnamefont
  {Molineux}}\ and\ \bibinfo {author} {\bibfnamefont {D.}~\bibnamefont
  {Panja}},\ }\href@noop {} {\bibfield  {journal} {\bibinfo  {journal} {Nature
  reviews. Microbiology}\ }\textbf {\bibinfo {volume} {11}},\ \bibinfo {pages}
  {194} (\bibinfo {year} {2013})}\BibitemShut {NoStop}%
\bibitem [{\citenamefont {Cordova}\ \emph {et~al.}(2003)\citenamefont
  {Cordova}, \citenamefont {Deserno}, \citenamefont {Gelbart},\ and\
  \citenamefont {Ben-Shaul}}]{cordova2003osmotic}%
  \BibitemOpen
  \bibfield  {author} {\bibinfo {author} {\bibfnamefont {A.}~\bibnamefont
  {Cordova}}, \bibinfo {author} {\bibfnamefont {M.}~\bibnamefont {Deserno}},
  \bibinfo {author} {\bibfnamefont {W.~M.}\ \bibnamefont {Gelbart}}, \ and\
  \bibinfo {author} {\bibfnamefont {A.}~\bibnamefont {Ben-Shaul}},\ }\href@noop
  {} {\bibfield  {journal} {\bibinfo  {journal} {Biophysical journal}\ }\textbf
  {\bibinfo {volume} {85}},\ \bibinfo {pages} {70} (\bibinfo {year}
  {2003})}\BibitemShut {NoStop}%
\bibitem [{\citenamefont {Grayson}\ \emph {et~al.}(2006)\citenamefont
  {Grayson}, \citenamefont {Evilevitch}, \citenamefont {Inamdar}, \citenamefont
  {Purohit}, \citenamefont {Gelbart}, \citenamefont {Knobler},\ and\
  \citenamefont {Phillips}}]{Grayson2006}%
  \BibitemOpen
  \bibfield  {author} {\bibinfo {author} {\bibfnamefont {P.}~\bibnamefont
  {Grayson}}, \bibinfo {author} {\bibfnamefont {A.}~\bibnamefont {Evilevitch}},
  \bibinfo {author} {\bibfnamefont {M.~M.}\ \bibnamefont {Inamdar}}, \bibinfo
  {author} {\bibfnamefont {P.~K.}\ \bibnamefont {Purohit}}, \bibinfo {author}
  {\bibfnamefont {W.~M.}\ \bibnamefont {Gelbart}}, \bibinfo {author}
  {\bibfnamefont {C.~M.}\ \bibnamefont {Knobler}}, \ and\ \bibinfo {author}
  {\bibfnamefont {R.}~\bibnamefont {Phillips}},\ }\href@noop {} {\bibfield
  {journal} {\bibinfo  {journal} {Virology}\ }\textbf {\bibinfo {volume}
  {348}},\ \bibinfo {pages} {430} (\bibinfo {year} {2006})}\BibitemShut
  {NoStop}%
\bibitem [{\citenamefont {Tzlil}\ \emph {et~al.}(2003)\citenamefont {Tzlil},
  \citenamefont {Kindt}, \citenamefont {Gelbart},\ and\ \citenamefont
  {Ben-Shaul}}]{Tzlil2003}%
  \BibitemOpen
  \bibfield  {author} {\bibinfo {author} {\bibfnamefont {S.}~\bibnamefont
  {Tzlil}}, \bibinfo {author} {\bibfnamefont {J.~T.}\ \bibnamefont {Kindt}},
  \bibinfo {author} {\bibfnamefont {W.~M.}\ \bibnamefont {Gelbart}}, \ and\
  \bibinfo {author} {\bibfnamefont {A.}~\bibnamefont {Ben-Shaul}},\ }\href@noop
  {} {\bibfield  {journal} {\bibinfo  {journal} {Biophysical journal}\ }\textbf
  {\bibinfo {volume} {84}},\ \bibinfo {pages} {1616} (\bibinfo {year}
  {2003})}\BibitemShut {NoStop}%
\bibitem [{\citenamefont {de~Gennes}\ and\ \citenamefont
  {Prost}(1993)}]{dGP93}%
  \BibitemOpen
  \bibfield  {author} {\bibinfo {author} {\bibfnamefont {P.~G.}\ \bibnamefont
  {de~Gennes}}\ and\ \bibinfo {author} {\bibfnamefont {J.}~\bibnamefont
  {Prost}},\ }\href@noop {} {\emph {\bibinfo {title} {The physics of liquid
  crystals}}}\ (\bibinfo  {publisher} {Oxford University Press},\ \bibinfo
  {year} {1993})\BibitemShut {NoStop}%
\bibitem [{\citenamefont {Kleman}\ and\ \citenamefont
  {Laverntovich}(2007)}]{kleman2007soft}%
  \BibitemOpen
  \bibfield  {author} {\bibinfo {author} {\bibfnamefont {M.}~\bibnamefont
  {Kleman}}\ and\ \bibinfo {author} {\bibfnamefont {O.~D.}\ \bibnamefont
  {Laverntovich}},\ }\href@noop {} {\emph {\bibinfo {title} {Soft matter
  physics: an introduction}}}\ (\bibinfo  {publisher} {Springer Science \&
  Business Media},\ \bibinfo {year} {2007})\BibitemShut {NoStop}%
\bibitem [{\citenamefont {Onsager}(1949)}]{onsager1949}%
  \BibitemOpen
  \bibfield  {author} {\bibinfo {author} {\bibfnamefont {L.}~\bibnamefont
  {Onsager}},\ }\href@noop {} {\bibfield  {journal} {\bibinfo  {journal}
  {Annals of the New York Academy of Sciences}\ }\textbf {\bibinfo {volume}
  {51}},\ \bibinfo {pages} {627} (\bibinfo {year} {1949})}\BibitemShut
  {NoStop}%
\bibitem [{\citenamefont {Doi}\ and\ \citenamefont
  {Kuzuu}(1985)}]{doi-kuzuu1985}%
  \BibitemOpen
  \bibfield  {author} {\bibinfo {author} {\bibfnamefont {M.}~\bibnamefont
  {Doi}}\ and\ \bibinfo {author} {\bibfnamefont {N.}~\bibnamefont {Kuzuu}},\
  }\href@noop {} {\bibfield  {journal} {\bibinfo  {journal} {J. Appl. Polym
  Sci: Appl. Polym. Symp.}\ }\textbf {\bibinfo {volume} {41}},\ \bibinfo
  {pages} {65} (\bibinfo {year} {1985})}\BibitemShut {NoStop}%
\bibitem [{\citenamefont {Purohit}\ \emph {et~al.}(2005)\citenamefont
  {Purohit}, \citenamefont {Inamdar}, \citenamefont {Grayson}, \citenamefont
  {Squires}, \citenamefont {Kondev},\ and\ \citenamefont
  {Phillips}}]{purohit2005forces}%
  \BibitemOpen
  \bibfield  {author} {\bibinfo {author} {\bibfnamefont {P.~K.}\ \bibnamefont
  {Purohit}}, \bibinfo {author} {\bibfnamefont {M.~M.}\ \bibnamefont
  {Inamdar}}, \bibinfo {author} {\bibfnamefont {P.~D.}\ \bibnamefont
  {Grayson}}, \bibinfo {author} {\bibfnamefont {T.~M.}\ \bibnamefont
  {Squires}}, \bibinfo {author} {\bibfnamefont {J.}~\bibnamefont {Kondev}}, \
  and\ \bibinfo {author} {\bibfnamefont {R.}~\bibnamefont {Phillips}},\
  }\href@noop {} {\bibfield  {journal} {\bibinfo  {journal} {Biophysical
  journal}\ }\textbf {\bibinfo {volume} {88}},\ \bibinfo {pages} {851}
  (\bibinfo {year} {2005})}\BibitemShut {NoStop}%
\bibitem [{\citenamefont {Ginzburg}\ and\ \citenamefont
  {Landau}(1950)}]{Landau1950}%
  \BibitemOpen
  \bibfield  {author} {\bibinfo {author} {\bibfnamefont {V.}~\bibnamefont
  {Ginzburg}}\ and\ \bibinfo {author} {\bibfnamefont {L.}~\bibnamefont
  {Landau}},\ }\href@noop {} {\bibfield  {journal} {\bibinfo  {journal} {Zh.
  Eksp. Teor. Fiz.}\ }\textbf {\bibinfo {volume} {20}},\ \bibinfo {pages} {164}
  (\bibinfo {year} {1950})}\BibitemShut {NoStop}%
\bibitem [{\citenamefont {Hardt}\ \emph {et~al.}(1987)\citenamefont {Hardt},
  \citenamefont {Kinderlehrer},\ and\ \citenamefont {Lin}}]{HaKiLi86}%
  \BibitemOpen
  \bibfield  {author} {\bibinfo {author} {\bibfnamefont {R.}~\bibnamefont
  {Hardt}}, \bibinfo {author} {\bibfnamefont {D.}~\bibnamefont {Kinderlehrer}},
  \ and\ \bibinfo {author} {\bibfnamefont {F.~H.}\ \bibnamefont {Lin}},\
  }\href@noop {} {\bibfield  {journal} {\bibinfo  {journal} {Comm. Math.
  Phys.}\ }\textbf {\bibinfo {volume} {105}},\ \bibinfo {pages} {547} (\bibinfo
  {year} {1987})}\BibitemShut {NoStop}%
\bibitem [{\citenamefont {Ball}\ and\ \citenamefont
  {Zarnescu}(2008)}]{ball2008orientable}%
  \BibitemOpen
  \bibfield  {author} {\bibinfo {author} {\bibfnamefont {J.~M.}\ \bibnamefont
  {Ball}}\ and\ \bibinfo {author} {\bibfnamefont {A.}~\bibnamefont
  {Zarnescu}},\ }\href@noop {} {\bibfield  {journal} {\bibinfo  {journal}
  {Molecular Crystals and Liquid Crystals}\ }\textbf {\bibinfo {volume}
  {495}},\ \bibinfo {pages} {221} (\bibinfo {year} {2008})}\BibitemShut
  {NoStop}%
\bibitem [{\citenamefont {Leiman}\ \emph {et~al.}(2003)\citenamefont {Leiman},
  \citenamefont {Kanamaru}, \citenamefont {Mesyanzhinov}, \citenamefont
  {Arisaka},\ and\ \citenamefont {Rossmann}}]{leiman2003structure}%
  \BibitemOpen
  \bibfield  {author} {\bibinfo {author} {\bibfnamefont {P.}~\bibnamefont
  {Leiman}}, \bibinfo {author} {\bibfnamefont {S.}~\bibnamefont {Kanamaru}},
  \bibinfo {author} {\bibfnamefont {V.}~\bibnamefont {Mesyanzhinov}}, \bibinfo
  {author} {\bibfnamefont {F.}~\bibnamefont {Arisaka}}, \ and\ \bibinfo
  {author} {\bibfnamefont {M.}~\bibnamefont {Rossmann}},\ }\href@noop {}
  {\bibfield  {journal} {\bibinfo  {journal} {Cellular and Molecular Life
  Sciences}\ }\textbf {\bibinfo {volume} {60}},\ \bibinfo {pages} {2356}
  (\bibinfo {year} {2003})}\BibitemShut {NoStop}%
\bibitem [{\citenamefont {Olson}\ \emph {et~al.}(2001)\citenamefont {Olson},
  \citenamefont {Gingery}, \citenamefont {Eiserling},\ and\ \citenamefont
  {Baker}}]{olson2001structure}%
  \BibitemOpen
  \bibfield  {author} {\bibinfo {author} {\bibfnamefont {N.~H.}\ \bibnamefont
  {Olson}}, \bibinfo {author} {\bibfnamefont {M.}~\bibnamefont {Gingery}},
  \bibinfo {author} {\bibfnamefont {F.~A.}\ \bibnamefont {Eiserling}}, \ and\
  \bibinfo {author} {\bibfnamefont {T.~S.}\ \bibnamefont {Baker}},\ }\href@noop
  {} {\bibfield  {journal} {\bibinfo  {journal} {Virology}\ }\textbf {\bibinfo
  {volume} {279}},\ \bibinfo {pages} {385} (\bibinfo {year}
  {2001})}\BibitemShut {NoStop}%
\bibitem [{\citenamefont {Bustamante}\ \emph {et~al.}(2003)\citenamefont
  {Bustamante}, \citenamefont {Bryant},\ and\ \citenamefont
  {Smith}}]{bustamante2003ten}%
  \BibitemOpen
  \bibfield  {author} {\bibinfo {author} {\bibfnamefont {C.}~\bibnamefont
  {Bustamante}}, \bibinfo {author} {\bibfnamefont {Z.}~\bibnamefont {Bryant}},
  \ and\ \bibinfo {author} {\bibfnamefont {S.~B.}\ \bibnamefont {Smith}},\
  }\href@noop {} {\bibfield  {journal} {\bibinfo  {journal} {Nature}\ }\textbf
  {\bibinfo {volume} {421}},\ \bibinfo {pages} {423} (\bibinfo {year}
  {2003})}\BibitemShut {NoStop}%
\bibitem [{\citenamefont {Kuzuu}\ and\ \citenamefont {Doi}(1984)}]{DK84}%
  \BibitemOpen
  \bibfield  {author} {\bibinfo {author} {\bibfnamefont {N.}~\bibnamefont
  {Kuzuu}}\ and\ \bibinfo {author} {\bibfnamefont {M.}~\bibnamefont {Doi}},\
  }\href@noop {} {\bibfield  {journal} {\bibinfo  {journal} {J. Phys. Soc.
  Japan}\ }\textbf {\bibinfo {volume} {53}},\ \bibinfo {pages} {1031} (\bibinfo
  {year} {1984})}\BibitemShut {NoStop}%
\bibitem [{\citenamefont {Wolfsheimer}\ \emph {et~al.}(2006)\citenamefont
  {Wolfsheimer}, \citenamefont {Tanase}, \citenamefont {Shundyak},
  \citenamefont {Van~Roij},\ and\ \citenamefont {Schilling}}]{vanRoij2005-1}%
  \BibitemOpen
  \bibfield  {author} {\bibinfo {author} {\bibfnamefont {S.}~\bibnamefont
  {Wolfsheimer}}, \bibinfo {author} {\bibfnamefont {C.}~\bibnamefont {Tanase}},
  \bibinfo {author} {\bibfnamefont {K.}~\bibnamefont {Shundyak}}, \bibinfo
  {author} {\bibfnamefont {R.}~\bibnamefont {Van~Roij}}, \ and\ \bibinfo
  {author} {\bibfnamefont {T.}~\bibnamefont {Schilling}},\ }\href@noop {}
  {\bibfield  {journal} {\bibinfo  {journal} {Physical Review E}\ }\textbf
  {\bibinfo {volume} {73}},\ \bibinfo {pages} {061703} (\bibinfo {year}
  {2006})}\BibitemShut {NoStop}%
\bibitem [{\citenamefont {van Roij}(2005)}]{vanRoij2005-2}%
  \BibitemOpen
  \bibfield  {author} {\bibinfo {author} {\bibfnamefont {R.}~\bibnamefont {van
  Roij}},\ }\href@noop {} {\bibfield  {journal} {\bibinfo  {journal} {European
  journal of physics}\ }\textbf {\bibinfo {volume} {26}},\ \bibinfo {pages}
  {S57} (\bibinfo {year} {2005})}\BibitemShut {NoStop}%
\bibitem [{\citenamefont {Priezjev}\ and\ \citenamefont
  {Pelcovits}(2000)}]{priezjev2000}%
  \BibitemOpen
  \bibfield  {author} {\bibinfo {author} {\bibfnamefont {N.}~\bibnamefont
  {Priezjev}}\ and\ \bibinfo {author} {\bibfnamefont {R.~A.}\ \bibnamefont
  {Pelcovits}},\ }\href@noop {} {\bibfield  {journal} {\bibinfo  {journal}
  {Physical Review E}\ }\textbf {\bibinfo {volume} {62}},\ \bibinfo {pages}
  {6734} (\bibinfo {year} {2000})}\BibitemShut {NoStop}%
\end{thebibliography}%
\end{document}